\documentclass[preprint,aps,pra,showpacs]{revtex4}
\usepackage{tabularx}
\usepackage{dcolumn}
\usepackage{graphics}% Include figure files
\usepackage{bm}% bold matha
\usepackage[dvips]{graphicx}
\usepackage[dvips]{rotating}
\usepackage[latin1]{inputenc}
\usepackage{dcolumn}
\usepackage{tabularx}
\usepackage{amsmath}
\usepackage{amssymb}
\begin{document}
\draft

\title{What is the difference in the p-wave and s-wave photodetachment in an electric field?}

\author{M. L. Du }\email{
duml@itp.ac.cn}
 \affiliation{Institute
of Theoretical Physics, Chinese Academy of Sciences, P. O. Box 2735,
Beijing 100190, China}

\date{\today}

\begin{abstract}
By applying closed-orbit theory to an existing model, a simple formula is derived for the modulation function of s-wave photo-detachment in the presence of a static electric field. We then compare the s-wave modulation function with the p-wave modulation function. We show the maximums  (minimums) in the s-wave modulation function correspond to the minimums (maximums) in the p-wave modulation function because of a phase difference of $\pi$ in their oscillations. The oscillation amplitude in the p-wave modulation function can be larger than, smaller than or equal to the oscillation amplitude of s-wave modulation function when the angle between the laser polarization direction and the static electric field direction is tuned to be smaller than, larger than or equal to a special angle which  is approximately $54.74^\circ$.

\par

\pacs{32.80.Gc,32.60.+i,31.15.Gy}

\end{abstract}

 \maketitle

Bryant \emph{et al}\cite{Bryant87,Bryant88} first observed the "ripple" structure in the photo-detachment cross section of H$^-$ in the presence of a static
electric field. The experimental observations have been successfully explained by several theories\cite{Fabrikant80,Rau88,Du88a}. The photodetachment of H$^-$ is an example of a p-wave photodetachment. It is well known that the initial bound state of H$^-$ posses an s symmetry. According to angular momentum selection rules, the photo-detached electron has a p-wave character in the absence of an electric field. Gibson \emph{et al} studied another p-wave photodetachment process in the presence of a static electric field using Au$^-$\cite{Gibson93b}. Gibson \emph{et al}\cite{Gibson93a,Gibson2001} also studied the oscillations in the photodetachment cross sections in the presence of an electric field using S$^-$ and Cl$^-$. However, these two negative ions are examples of s-wave photodetachment. In contrast to p-wave photodetachment, detached electron wave in an s-wave photodetachment has an s-wave character. In reality, s-wave photo-detachment is an approximation. Because the detached electron should be a combination of s-wave component and d-wave component according to selection rules. But close to threshold, the d-wave is much smaller than the s-wave according to Wigner's threshold law and can be neglected. Since the initial bound states of negative ions have different symmetries for s-wave and p-wave photo-detachment, the question is how the symmetry  properties in the initial bound states are reflected in the static field induced oscillations of the photo-detachment cross sections? So far, comparisons between theoretical calculations and experimental measurements have been limited to each of the negative ions such as H$^-$,Au$^-$,S$^-$ and Cl$^-$\cite{Bryant87,Bryant88,Du88a,Gibson93b,Gibson93a,Gibson2001}. There has been no comparison between the oscillations of different negative ions. The lack of such comparison is probably caused by the difficulty to identify a simple and common object characterizing the oscillations of each negative ion.

It appears that modulation functions as functions of a scale variable $S$ are the choice to compare the oscillations of different negative ions. For H$^-$ photodetachment in an electric field, both momentum space approach\cite{Du88a} and closed-orbit theory\cite{Du04} give the following photodetachment cross section when the laser polarization is parallel to the electric field,
\begin{equation}
\sigma_{||}(E,F)=\sigma_0(E)[1+\frac{\cos(S)}{S}],
\end{equation}
where $\sigma_0(E)$ is the photodetachment cross section of H$^-$ in the absence of a static electric field, $E$ is the detached electron energy above threshold, $S=\frac{4\sqrt{2}E^{3/2}}{3F}$ is the action of a closed-orbit of the detached-electron\cite{Du04} that is responsible for  the oscillation in Eq.(1), $F$ is the static electric field strength.
The induced oscillation by the static electric field in the photodetachment cross section for H$^-$ can be characterized by a  modulation function,
\begin{equation}
A^{||}_p(S)=[1+\frac{\cos(S)}{S}],
\end{equation}
which is the ratio of the photodetachment cross section in the static electric field and the photodetachment cross section in the absence of the static electric field.

For the purpose of comparison, we now derive a similar modulation function for s-wave photodetachment in an electric field using closed-orbit theory\cite{Du04}. Following Baruch \emph{et al}\cite{Maruch,Gibson93a,Gibson2001}, we assume the dipole operator acting on the initial state can be approximated by a $\delta$ source $D(\textbf{r})\psi_i(\textbf{r})=\lambda \delta(\textbf{r})$ for s-wave photodetachment, where $\lambda$ is an energy dependent positive parameter which will be related to the photodetachment cross section in the absence of a static electric field.
The photo-detached electron wave function $\psi_s^+(\textbf{r})$ in the presence of an electric field of $F$ pointing to the z-axis satisfies the following equation\cite{Du04,Du87,Du88b},
\begin{equation}
(E-H)\psi_s^{+}(\textbf{r})=\lambda \delta(\textbf{r}),
\end{equation}
where $H=\frac{\textbf{p}^2}{2}+Fz$. Both experimental and theoretical evidences suggest the re-scattering effect is small\cite{Gibson2001,Fabrikant89,Fabrikant94} and is neglected here.
The physical solution of Eq.(3) must be outgoing.
Once we have the solution $\psi_s^+(\textbf{r})$, the photodetachment cross section can be calculated using the following formula\cite{Du04,Du87,Du88b}
\begin{equation}
\sigma_s(E,F)=-\frac{4\pi E_p}{c}Im\langle \lambda \delta(\textbf{r})|\psi_s^+(\textbf{r})\rangle,
\end{equation}
where $E_p$ is the photon energy, and $c$ is approximately equal to 137. We use atomic units in this article.
Following previous approach for H$^-$\cite{Du04},
the photodetachment wave function $\psi_s^+(\textbf{r})$ is written as a direct part and a returning part $\psi_s^+(\textbf{r})=\psi^+_{s,dir}(\textbf{r})+\psi^+_{s,ret}(\textbf{r})$.
The direct part represents the detached electron wave initially going out from the source right after photodetachment and it satisfies the following equation
\begin{equation}
(E-\frac{\textbf{p}^2}{2})\psi^+_{s,dir}(\textbf{r})=\lambda \delta(\textbf{r}).
\end{equation}
Using existing result for Green's function\cite{Jackson}, we have
\begin{equation}
\psi^+_{s,dir}(\textbf{r})=-\frac{\lambda}{2\pi}\frac{\exp(ikr)}{r}
\end{equation}
where $r$ is the distance to the delta function source at the origin,
$k=\sqrt{2E}$.

To obtain the returning wave $\psi^+_{s,ret}(\textbf{r})$, we propagate the direct wave in Eq.(6) using semiclassical method following the same closed-orbit\cite{Du04}. The returning wave near the source can be approximated by a plane wave traveling in the negative z-direction, but its phase and amplitude are different from the ones for H$^-$\cite{Du04},
\begin{equation}
\psi^+_{s,ret}(\textbf{r})=g_s \exp(-ikz),
\end{equation}
where $g_s$ carries the amplitude and phase accumulated along the closed-orbit and the direct wave function $\psi^+_{s,dir}(\textbf{r})$. According to the semiclassical method\cite{Du04},
\begin{equation}
g_s=\lim_{R\rightarrow 0}Ae^{i(S-\frac{\pi}{2})}\psi^+_{s,dir}(R).
\end{equation}
In Eq.(8) $A$ is a measure of the amplitude variation along the closed-orbit and is given by
\begin{equation}
A=\sqrt{\frac{R^2k}{(R+kT)^2|k-FT|}},
\end{equation}
where $T$ is the transit time of the closed-orbit and is given by  $T=\frac{2\sqrt{2E}}{F}$. $S$ is the action integral along the closed-orbit from the source out and back to the source,
$S(E,F)=\frac{4\sqrt{2}E^{3/2}}{3F}$. For this closed-orbit, the general relationship $T=\frac{\partial S}{\partial E}$ obviously holds\cite{Du88b}.

Finally we get the result
\begin{equation}
g_s=-\frac{\lambda F}{4\pi k^2}e^{i(S-\frac{\pi}{2})}.
\end{equation}

The photodetachment cross section can now be evaluated using Eq.(4). The cross section can be written as
\begin{equation}
\sigma_s(E,F)=\sigma_{s,0}(E)+\sigma_{s,1}(E,F)
\end{equation}
where $\sigma_{s,0}(E)$ and $\sigma_{s,1}(E,F)$ are,respectively,
\begin{eqnarray}
\sigma_{s,0}(E,F) & = & -\frac{4\pi E_p}{c}Im\langle \lambda\delta(\textbf{r})|\psi^+_{s,dir}(\textbf{r})\rangle,\\
\sigma_{s,1}(E,F) & = & -\frac{4\pi E_p}{c}Im\langle \lambda\delta(\textbf{r})|\psi^+_{s,ret}(\textbf{r})\rangle,
\end{eqnarray}

We evaluate the background term $\sigma_{s,0}(E)$ first.
\begin{eqnarray}
\sigma_{s,0}(E) & = & -\frac{4\pi E_p}{c}\Im\langle \lambda\delta(\textbf{r})|\psi^+_{dir}(\textbf{r})\rangle\nonumber\\
 & = & -\frac{4\pi E_p}{c}\Im \int d\textbf{r} \lambda\delta(\textbf{r})(-\frac{\lambda}{2\pi})\frac{\exp(ikr)}{r}\nonumber\\
 & = & -\frac{4\pi E_p}{c}\int d\textbf{r} \lambda\delta(\textbf{r})(-\frac{\lambda}{2\pi})\frac{\sin(kr)}{r}\nonumber\\
 & = & -\frac{4\pi E_p}{c}(-\frac{\lambda^2}{2\pi})k \nonumber\\
 & = & \frac{2\lambda^2E_pk}{c}.
\end{eqnarray}
The oscillating term $\sigma_{s,1}(E,F)$ is evaluated next.
\begin{eqnarray}
\sigma_{s,1}(E,F) & = & -\frac{4\pi E_p}{c}\Im\langle \lambda\delta(\textbf{r})|\psi^+_{ret}(\textbf{r})\rangle\nonumber\\
 & = & -\frac{4\pi E_p}{c} \Im \int d\textbf{r} \lambda\delta(\textbf{r})(-\frac{\lambda F}{4\pi k^2})\exp[i(S-\frac{\pi}{2})]\exp(-ikz)\nonumber\\
  & = & -\frac{4\pi E_p}{c}(-\frac{\lambda^2F}{4\pi k^2}) \Im \exp[i(S-\frac{\pi}{2})] \nonumber\\
 & = & -\frac{4\pi E_p}{c}(-\frac{\lambda^2F}{4\pi k^2})\sin(S-\frac{\pi}{2})\nonumber\\
 & = & \frac{\lambda^2E_pF}{k^2c}\sin(S-\frac{\pi}{2})\nonumber\\
 & = & -\frac{\lambda^2E_pF}{k^2c}\cos(S).
\end{eqnarray}
Combining the above two terms, we finally have the cross section of s-wave photodetachment in an electric field,
\begin{equation}
\sigma_s(E,F)=\sigma_{s,0}(E)[1-\frac{\cos(S)}{3S}]
\end{equation}
$\sigma_{s,0}(E)=\frac{2\lambda^2E_pk}{c}$ is the photodetachment cross section without the static electric field. The energy dependent parameter $\lambda$ is thus connected to $\sigma_{s,0}(E)$. We note the $\delta$ source model gives the correct threshold law for s-wave photodetachment. $\lambda$ varies slowly and smoothly as a function of energy above and near threshold. The modulation function for s-wave photodatachment in the presence of a static electric field is therefore given by
\begin{equation}
A_s(S)=[1-\frac{\cos(S)}{3S}].
\end{equation}
The modulation function is expected to be valid for $S\geq1$ because of the semiclassical approximation in the derivation.

We now compare the modulation functions of s-wave in an electric field in Eq.(17) and p-wave photodetachment in an electric field in the parallel  polarization case in Eq.(1). First, we note both the s-wave and p-wave modulation functions are functions of the same closed-orbit action $S$, which in turn depends on the detached electron energy $E$ and static electric field $F$ via $S(E,F)=\frac{4\sqrt{2}E^{3/2}}{3F}$. This  observation facilities the comparison of the modulation functions of s-wave and p-wave. As functions of $S$, we realize that the phases of the oscillations in Eq.(17) and Eq.(1) differ by $\pi$. This phase difference implies the maximums (minimums) of s-wave modulation function correspond to the minimums (maximums) of p-wave modulation function. In Fig.1(a) the s-wave modulation function $A_s(S)$ (solid line) and $A^{||}_p(S)$ (dotted line) are compared. In Fig.1(a) one can also see the oscillation amplitude of p-wave in the parallel polarization case is larger than the oscillation amplitude of s-wave. In fact, Eq.(17) shows the oscillation amplitude of p-wave in the parallel polarization case is three times of the oscillation amplitude of s-wave oscillation.

The discussion for the p-wave can be extended to more general situations.  Let $\theta_L$ be the angle between the laser polarization direction and the static electric field. When $\theta_L$ is not too close to $\pi/2$, the photodetachment cross section derived earlier\cite{Du06} gives the   $\theta_L$ dependent modulation function of p-wave,
\begin{equation}
A_p(S,\theta_L)=[1+\cos^2(\theta_L)\frac{\cos(S)}{S}].
\end{equation}
When $\theta_L$ is very close to $\pi/2$, the outgoing detached electron wave is on a node, the result in Eq.(18) will be modified similar to the photo-ionization case\cite{node}.

The phase of the oscillation in Eq.(18) remains the same as in the parallel polarization case in Eq.(1) when $\theta_L$ deviates from zero. However, the oscillation amplitude depends on the angle $\theta_L$ and is reduced by a factor $\cos^2(\theta_L)$. By setting $\theta_L$ to a special angle
satisfying $\cos(\theta^s_L)=\frac{1}{\sqrt{3}}$ the oscillation amplitudes in the modulation functions of p-wave and s-wave will be the same. Limiting the value of $\theta_L$ from $0^\circ$ to $90^\circ$, the special angle is found to be  $\theta^s_L\doteq54.74^\circ$.
In Fig.1(b) we compare the modulation functions of s-wave and p-wave for this special situation. The solid curve represents the modulation of s-wave and the dotted curve represents the modulation of p-wave. The oscillating parts are mirror images of each other with respect to the horizontal line intersecting the vertical axis at 1. $\theta^s_L$ is the dividing point in the range [$0^\circ$,$90^\circ$]. When $\theta_L<\theta^s_L$, the oscillation amplitude of the p-wave is larger than the oscillation amplitude of the s-wave; when $\theta_L>\theta^s_L$, the oscillation amplitude of the p-wave is smaller the oscillation amplitude of the s-wave. So far, experimental measurements for p-wave photo-detachment have been reported for the parallel polarization case corresponding to $\theta_L=0^\circ$ and for the perpendicular case corresponding to $\theta_L=90^\circ$\cite{Bryant87,Bryant88,Gibson93b}. The present results suggest the oscillation of the p-wave modulation function at other angles between the laser polarization and static electric field can provide useful informations.

In summary, we have derived a modulation function for s-wave photo-detachment in a static electric field in Eq.(17). When the semiclassical condition is satisfied, that is, when  $S(E,F)=\frac{4\sqrt{2}E^{3/2}}{3F}\geq1$, both the s-wave and p-wave modulation functions depend on only the closed-orbit action $S$.
We have compared the s-wave and p-wave modulation functions as functions of $S$. We find the phase of the oscillation in the s-wave modulation function differs by $\pi$ from the phase of the oscillation in the p-wave modulation. The origin of this phase difference can be traced back to the difference in the symmetry properties of the initial bound states of the negative ions. The oscillation amplitude of p-wave modulation function can be made larger than, smaller than or equal to the oscillation amplitude of s-wave modulation function when the angle between the laser polarization and the static electric field is tuned to be smaller than, larger than or equal to a special angle $\theta^s_L$ which is approximately $54.74^\circ$. We hope the present theoretical results will provide useful guide for future experiments.

%\newpage

%\end{multicols}

\newpage
\begin{figure}

\includegraphics[scale=1.0,angle=-0]{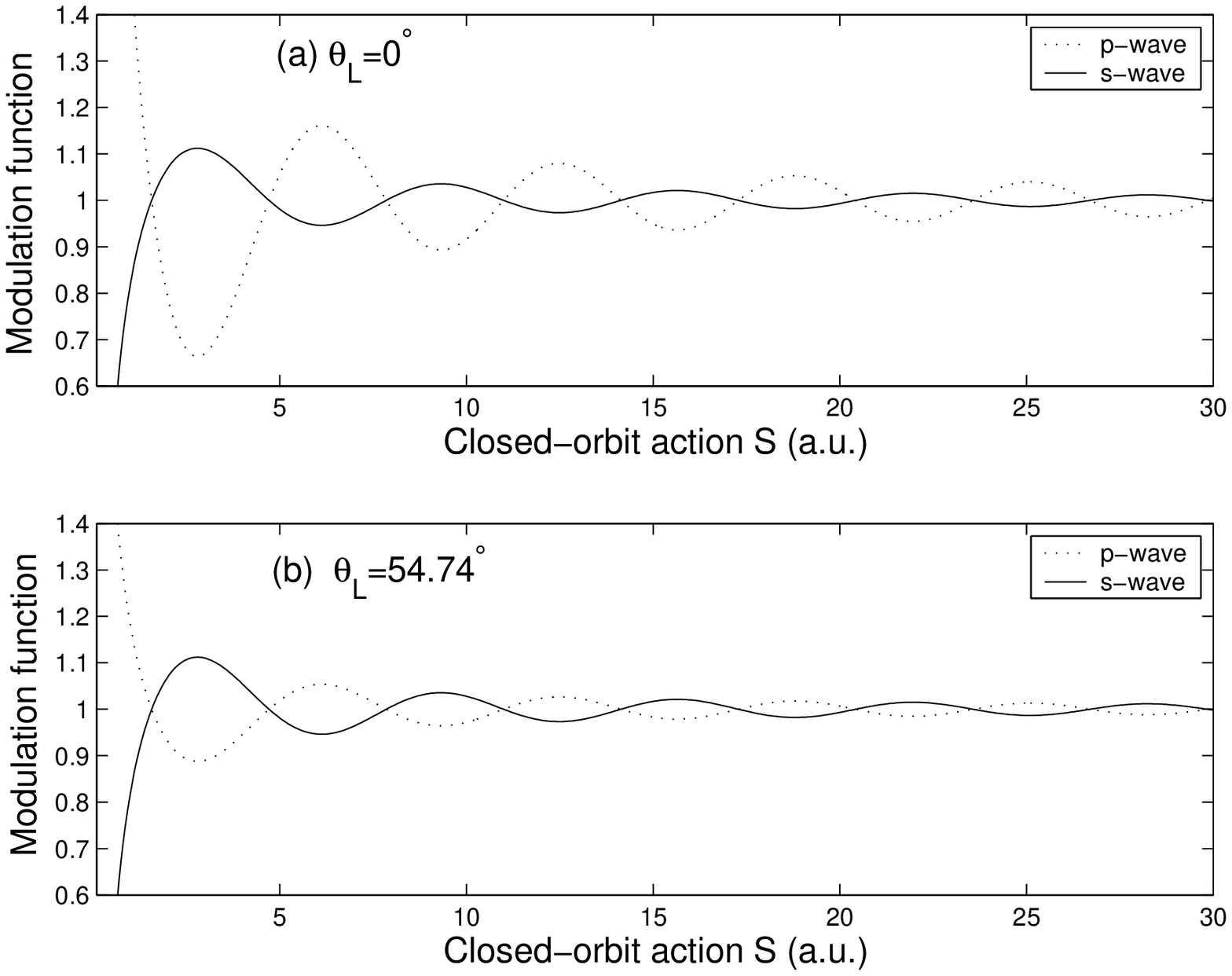}
\caption{Modulation functions of s-wave and p-wave photo-detachment in the presence of a static electric field. (a) The angle between the laser polarization direction and the static electric field $\theta_L$ is $0^\circ$. (b) The angle between the laser polarization direction and the static electric field $\theta_L$ is $54.74^\circ$.
  }
\end{figure}

%\end{multicols}{2}


\begin{references}



\bibitem{Bryant87}H.C.Bryant,A.Mohagheghi,J.E.Stewart,
      J.Donahue,C.R.Quick,
      R.A.Reeder,\\V.Yuan,C.Hummer,W.W.Smith,S.Cohen,et
      al.,Phys.Rev.Lett.{\bf 58},2412 (1987).
\bibitem{Bryant88}J.E.Stewart,H.C.Bryant,P.G.Harris,A.H.Mohagheghi,
      J.B.Donahue,C.R.Quick,\\R.A.Reeder,V.Yuan,
      C.R.hummer,W.W.Smith,
      et al.,Phys.Rev.A {\bf 38},5628 (1988).
\bibitem{Fabrikant80}I.I. Fabrikant, Sov. Phys. JETP 52, 1045 (1980).
\bibitem{Rau88}A.R.P.Rau and H.Wong,Phys.Rev.A {\bf 37},632 (1988).
\bibitem{Du88a}M.L.Du and J.B.Delos,Phys.Rev.A {\bf 38},5609 (1988).
\bibitem{Gibson93b}N.D.Gibson,B.J.Davis,and D.J.Larson, Phys.Rev.A {\bf
      48},310 (1993).
\bibitem{Gibson93a}N.D.Gibson,B.J.Davis,and D.J.Larson, Phys.Rev.A {\bf
      47},1946(1993).
\bibitem{Gibson2001} N. D. Gibson, M. D. Gasda, K. A. Moore, D. A. Zawistowski, and C. W. Walter, Phys.Rev.A  {\bf
      64},061403 (2001).
\bibitem{Du04}M.L.Du,Phys.Rev.A {\bf70},055402 (2004).
\bibitem{Maruch}M. C. Baruch, W. G. Sturrus, N. D. Gibson, and D. J. Larson, Phys.Rev.A {\bf45},2825 (1992).
\bibitem{Du87}M.L.Du and J.B.Delos,Phys.Rev.Lett. {\bf 58},1731 (1987).
\bibitem{Du88b}M.L.Du and J.B.Delos,Phys.Rev.A {\bf 38},1896,1913 (1988).
\bibitem{Fabrikant89}I. I. Fabrikant, Phys. Rev. A {\bf 40}, 2373 (1989).
\bibitem{Fabrikant94}I. I. Fabrikant, J. Phys. B {\bf 27}, 4545 (1994).
\bibitem{Jackson}J. D. Jackson, Classical Electrodynamics, Wiley Press, New York (1975),p.224.
\bibitem{Du06}M.L.Du,Eur.Phys.J.D. {\bf38},533 (2006).
\bibitem{node}John A. Shaw, John B. Delos, Michael Courtney, and Daniel Kleppner,Physical Review A {\bf 52}, 3695 (1995).



\end{references}
\end{document}